\documentclass[useAMS,usenatbib]{mn2e} \usepackage{graphicx}
\title[A Multibeam Survey of the Galaxy for Methanol Masers]{The 6-GHz
  Multibeam Maser Survey I. Techniques}

\author[Green et al.]
       {J. A.
Green$^1$\thanks{E-mail:james.green@csiro.au}, J. L.
Caswell$^2$, G. A. Fuller$^1$, A. Avison$^{1}$, S. L. Breen$^{2,3}$, K. Brooks$^2$, \newauthor
       M. G. Burton$^4$, A. Chrysostomou$^5$, J. Cox$^6$, P. J.
Diamond$^1$, S. P. Ellingsen$^3$, \newauthor
       M. D. Gray$^1$, M. G. Hoare$^7$, M. R. W. Masheder$^8$, N. M.
McClure-Griffiths$^2$, \newauthor
       M. Pestalozzi$^{5,11}$, C. Phillips$^2$, L. Quinn$^1$, M. A.
Thompson$^5$, M. A. Voronkov$^2$, \newauthor
       A. Walsh$^{9}$, D. Ward-Thompson$^6$, D. Wong-McSweeney$^1$, J. A.
Yates$^{10}$ \newauthor
        and R. J. Cohen$^{1}$\thanks{Deceased 2006 November 1.} \\ $^{1}$
Jodrell Bank Centre for Astrophysics, Alan Turing Building, University of
Manchester, Manchester, M13 9PL, UK; \\ $^{2}$ Australia Telescope
National Facility, CSIRO, PO Box 76, Epping, NSW 2121, Australia; \\
$^{3}$ School of Mathematics and Physics, University of Tasmania, Private
Bag 37, Hobart, TAS 7001, Australia; \\ $^{4}$ School of Physics,
University of New South Wales, Sydney, NSW 2052, Australia;\\ $^{5}$
Centre for Astrophysics Research, Science and Technology Research
Institute, University of Hertfordshire, College Lane, \\Hatfield, AL10
9AB, UK; \\ $^{6}$ Department of Physics and Astronomy, Cardiff
University, 5 The Parade, Cardiff, CF24 3YB, UK; \\ $^{7}$ School of
Physics and Astronomy, University of Leeds, Leeds, LS2 9JT, UK; \\ $^{8}$
Astrophysics Group, Department of Physics, Bristol University, Tyndall
Avenue, Bristol, BS8 1TL, UK;  \\ $^{9}$ School of Maths, Physics and IT,
James Cook University, Townsville, QLD 4811, Australia;\\ $^{10}$
University College London, Department of Physics and Astronomy, Gower
Street, London, WC1E 6BT, UK\\ $^{11}$ G\"oteborgs Universitet
Insitutionen f\"or Fysik, G\"oteborg, Sweden \\}

\date{Accepted XXXX . Received XXX; in original form XXXX}

\pagerange{\pageref{firstpage}--\pageref{lastpage}} \pubyear{XXXX}
\begin{document} \maketitle

\label{firstpage}

\begin{abstract}
A new 7-beam 6$-$7 GHz receiver has been built to survey
the Galaxy and the Magellanic Clouds for newly forming high-mass
stars that are pinpointed by strong methanol maser emission at 6668
MHz. The receiver was jointly constructed by Jodrell Bank Observatory
(JBO) and the Australia Telescope National Facility (ATNF) and allows 
simultaneous coverage at 6668 and 6035 MHz.  It was 
successfully commissioned at Parkes in January 2006 and is now being 
used to conduct the Parkes-Jodrell multibeam maser survey of the Milky 
Way.  This will be the first systematic survey of the entire 
Galactic plane for masers of not only 6668-MHz 
methanol, but also 6035-MHz excited-state hydroxyl.  The survey is two 
orders of magnitude faster than most previous systematic surveys and has
an rms noise level of $\sim$0.17\,Jy.
This paper describes the observational strategy, techniques 
and reduction procedures of
  the Galactic and Magellanic Cloud surveys, together with  deeper, pointed,
  follow-up observations and complementary observations with other 
instruments. It also includes an estimate of the survey detection efficiency.
  The 111 days of observations with the Parkes telescope have
  so far yielded $>$800 methanol sources, of which $\sim$350 are new
  discoveries.  The whole project will provide the first comprehensive Galaxy-wide
 catalogue of 6668-MHz and 6035-MHz masers.
\end{abstract}


\section{Introduction}

The theory and understanding of the formation of high-mass stars is a
fundamental unsolved problem in astronomy.  Current star formation models
encounter difficulties in creating stars much more massive than $\sim$8
M$_\odot$ \citep[e.g.][]{keto03, yorke04} and observations of these events
are hampered by the speed with which the process occurs, and hence the
relative rarity of objects in the act of accreting the majority of their
final mass.  However, since its discovery \citep{menten91}, the 6668-MHz
methanol maser has been recognised as one of the clearest signposts to
the formation of high-mass young stars.  It is the second strongest cosmic
maser known, surpassed only by water (H$_{2}$O) at 22~GHz. The 6668-MHz
methanol maser is widespread and, unlike all other strong masers (OH,
H$_{2}$O and SiO), is only found close to high-mass young stars \citep[e.g.][]{minier03}
and sometimes associated with ultra compact (UC) H{\sc ii}
regions. Additionally, the methanol masers may trace earlier hot core phases
\citep[e.g.][]{minier05, purcell06}, although recent kinematic work by \citet{walt07} argues otherwise.  
It is possible that heating of the hot
molecular core, or shock heating from an outflow liberates methanol from
dust mantles and so provides the high column density needed to give strong
maser action \citep{cragg02,codella04}.

\begin{figure*} 
\begin{center} 
\includegraphics[width=16cm]{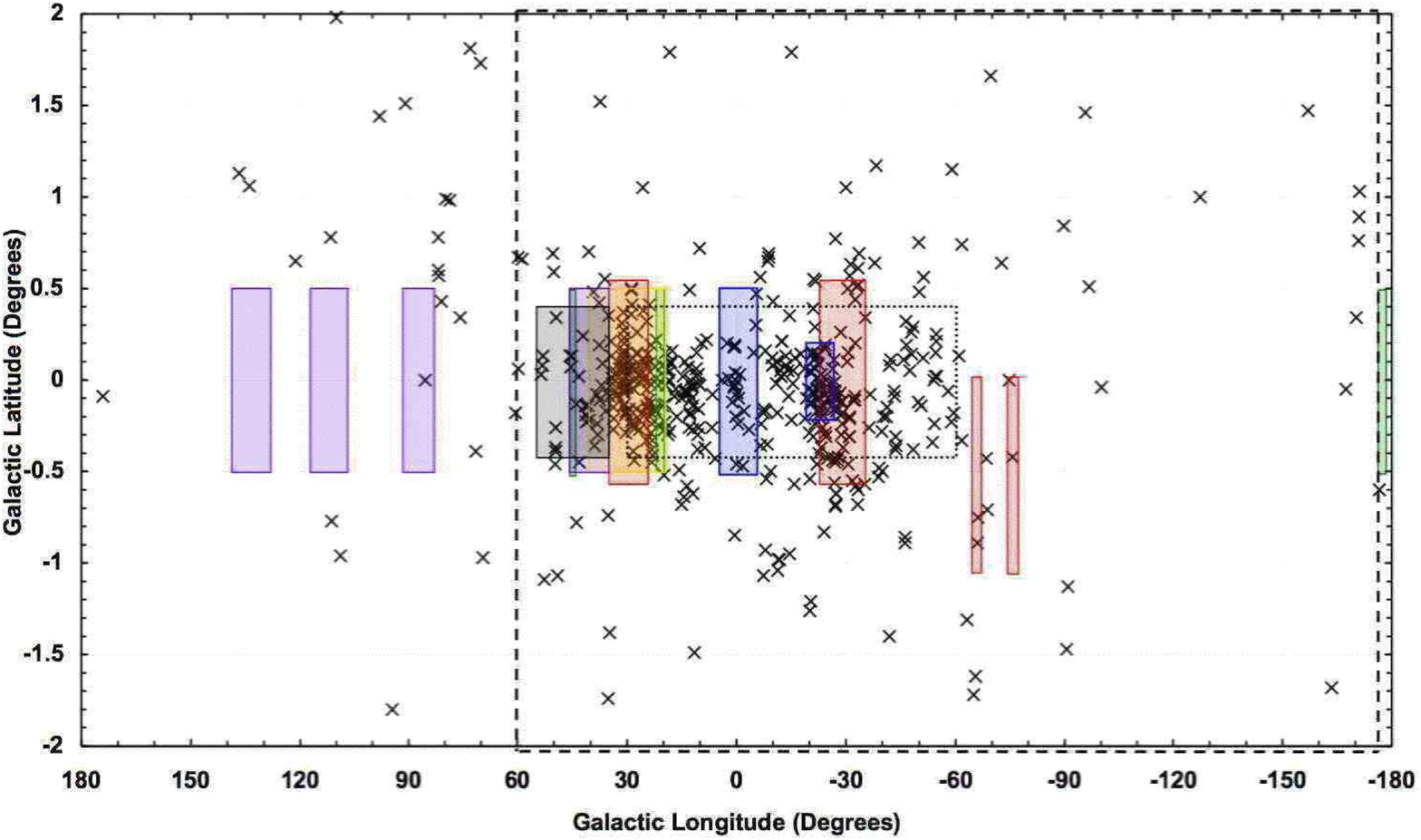}
\end{center} 
\caption{Galactic population of 6668-MHz methanol masers
within $\mid b \mid \le 2^{\circ}$, from \citet{pestalozzi05} catalogue. 
Boxed regions highlight previous surveys with the online version having
blue boxes represent the unbiased surveys of
\citet{caswell96a,caswell96b}; green \citep[private communication, see
][]{pestalozzi05};  yellow \citep{szymczak02}; red \citep{ellingsen96};
purple \citep{pestalozzi02a}; black \citep{pandian07a}. Dashed
line delineates the Parkes MMB survey region. Dotted line delineates the main area of the `Piggyback' survey (\S \ref{PuPiPr}).} 
\label{GALPOP} 
\end{figure*}

During the past decade, extensive 6668-MHz methanol maser searches have
been undertaken in our Galaxy using two different strategies: (1) targeted
searches toward colour--selected infrared (IR) sources and known regions
of intense star formation (e.g. OH and H$_2$O masers); (2)
unbiased surveys covering portions of the Galactic plane in both the
northern and southern hemispheres. These methods have yielded different
results in terms of detection statistics.  Except for the first targeted
survey by \citet{menten91} and the survey of OH maser positions by
\citet{caswell95a} (with success rates exceeding 80 per cent, and 
further augmented by the discovery of offset and clustered maser
sites), other targeted methanol maser
searches have produced detection rates of about 15 per cent
\citep{macleod92a,macleod92b,gaylard93,schutte93,walt95,walt96,walsh97, 
slysh99,szymczak00,ellingsen07}.   Generally, the targeted surveys were 
restricted to known 
or suspected sites of star formation and missed masers arising in regions
where they were not expected.  The unbiased searches have
produced detection rates lying between 0.22 sources degree$^{-2}$
\citep*{pestalozzi05} and more than 10 sources degree$^{-2}$ \citep{caswell96a,
caswell96c, ellingsen96}.  A total of 140 degree$^2$ have been
surveyed to date. This includes a region of 16 degree$^2$ in the first Galactic quadrant 
which have been covered in several unbiased surveys (as evident in Table 3 of \citealt{pestalozzi05}).
More than 500 sources are
listed in the compilation of 6668-MHz methanol sources by
\citet{pestalozzi05}. Fig.\,\ref{GALPOP} shows how these masers are
distributed in the Galaxy (for $\mid b \mid \le 2.0^{\circ}$).

Almost 50 per cent of the detected objects of
the untargeted surveys of \citet{ellingsen96} and \citet{szymczak02} had no 
previously known counterpart, clearly showing the limitation
of targeted observations. In both samples, possible counterparts to the masers have a 
wide range of infrared colours, with many conspicuously falling outside 
the UC H{\sc ii} region {\it IRAS} selection criteria of \citet{wood89}. 
Therefore, any survey based solely upon long-wavelength infrared colour 
selection criteria is likely to underestimate the number of masers, and searches using targets based on 
mid-infrared colour-selected sources of the {\it Spitzer} Galactic Legacy Infrared Mid-Plane Survey Extraordinaire (GLIMPSE) infrared survey  
\citep{ellingsen07}  also seem to be inefficient. 

On the other hand, the existing unbiased surveys are deficient because of 
their incomplete Galactic coverage and heterogeneous sensitivity limits.
A recent analysis of the existing surveys suggests that the actual number
of methanol masers sources in our Galaxy is about twice that currently
known \citep{walt05, pandian07b}.  A potential doubling of the number of known young
high mass stars clearly has important implications for our view of high-mass
star formation in our Galaxy, and provides the opportunity of enhancing our 
understanding of the nature and origin of 6668-MHz methanol masers.

The only way to determine the actual number of methanol maser sources in
the Galaxy, and identify the high-mass star formation regions which they
trace, is to carry out a uniform, unbiased survey of the whole Galactic
plane. With sufficient sensitivity, such a survey will detect
the vast majority of such sources.  Since the high brightness of masers
allows their positions and velocities to be determined to high precision,
such a Galaxy-wide survey will provide an excellent probe of the global
distribution of star formation throughout the Galaxy. In addition, the
unbiased, homogenous catalogue that such a survey can produce will 
provide a
statistically complete sample for future studies of the early stages of
high-mass star formation.

\section{The MMB Galactic Plane Survey} 
The need for a new survey of the 
Galactic Plane for the 6668-MHz methanol masers led
to the development of the Methanol Multibeam (MMB) survey. The survey 
goal is to produce a high-reliability unbiased survey of 6668-MHz methanol masers 
in the Galactic plane, with a uniform sensitivity completeness.  
It will allow a complete census of the high-mass star formation taking 
place in
our Galaxy.  In this way, and in combination with other Galaxy-wide
surveys, it will be possible to accumulate the statistics necessary to
determine the relative lifetimes of the various stages of high-mass star formation,
and hence be able to build an empirical model of the evolutionary process.
This paper and its forerunner \citep{green08a} represent the first two
in a series on the MMB survey, soon to be followed by the first Galactic results.

\subsection{Survey Parameters} 
To cover the whole Galactic plane in
longitude the MMB team has used the Parkes 64-m radio telescope in the 
southern hemisphere and currently plans to use the Lovell 76-m radio telescope at 
Jodrell Bank in the northern hemisphere. The survey area has been split, 
with longitudes $-$174$^{\circ}$ $< l <$ 60$^{\circ}$ observed from Parkes and 60$^{\circ}$ $< l <$ 186$^{\circ}$ 
planned for the northern hemisphere observations.

\begin{figure} \includegraphics[width=8.1cm]{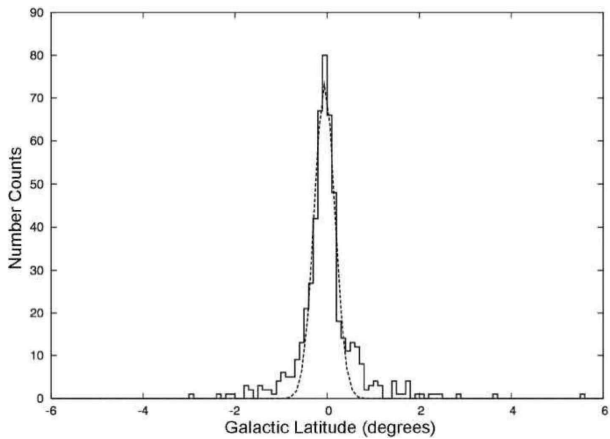}
\caption{Latitude
distribution of more than 500 methanol masers listed in the \citet{pestalozzi05}
compilation. Dotted line is a Gaussian fit with a FWHM of 0.52 degrees.} \label{Lat3} 
\end{figure}

The latitude distribution of methanol masers in the \citet{pestalozzi05}
compilation has a FWHM of $\sim$0.5$^{\circ}$ (Fig.\,\ref{Lat3}), but shows
quite broad wings. Also, there is the warp of the Galactic plane, clearly delineated 
by HI and continuum observations \citep[e.g.][]{burton76, dickey90, russeil03}. The importance of the warp has been reinforced by the 
latest molecular cloud surveys which reveal CO clouds in the outer Galaxy more
than 1\,kpc from the midplane \citep[e.g.][]{mizuno04}.  Furthermore, Infrared
Dark Clouds (IRDCs) detected by the Midcourse Space Experiment (MSX), and recognised as potential star
formation regions in their early stages \citep{carey00}, have up to 50 per
cent of their population outside $\mid b \mid \le 0.5^{\circ}$. 
Taking these factors into account and balancing
them against feasible observing time-scales, we adopted a
latitude range of $\mid b \mid \le 2^{\circ}$.  For the
\citet{pestalozzi05} compilation 14 sources lie outside this region
(four in the inner Galaxy, within 60 degrees of the Galactic centre, and 
ten in the outer Galaxy).

\begin{figure*} \begin{center}
\includegraphics[width=17.2cm]{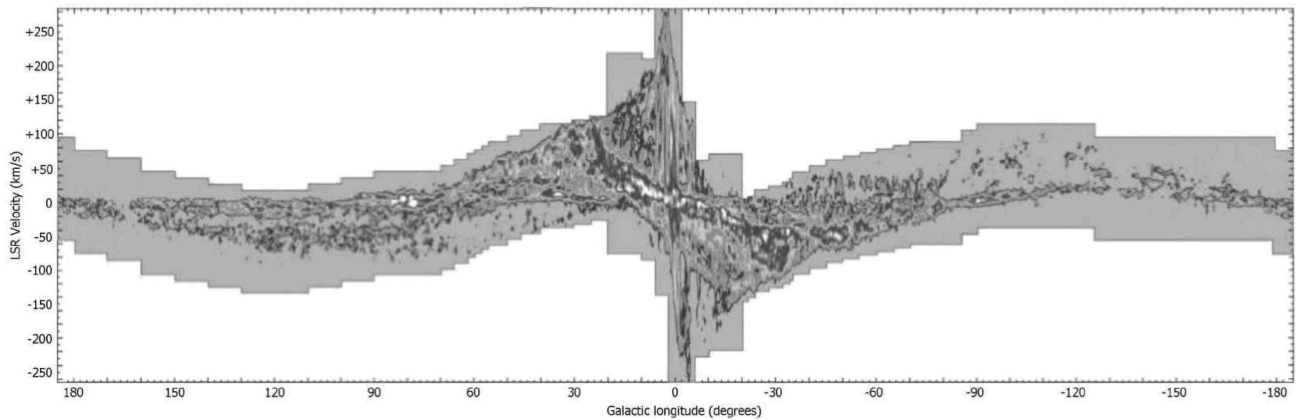} 
\end{center}
\caption{Longitude versus LSR velocity (km\,s$^{-1}$) plot of the greyscale of CO $J = 1
- 0$ from \citet{dame01} with grey shading
delineating the velocity coverage of the MMB survey.} 
\label{VELDIST}
\end{figure*}

The velocities of the molecular clouds in which the 6668-MHz methanol
masers reside are well traced by the emission from the CO molecule.  Thus 
fully sampling the observed velocity spread of CO emission is expected to 
adequately cover the range of velocities exhibited by 6668-MHz methanol 
masers (Section \ref{dataacq}). This velocity coverage is a function of Galactic 
longitude and is shown on Fig.\,\ref{VELDIST} overlaid on the 
position-velocity plot of the CO emission from \citet{dame01}.

The targeted 1$\sigma$ noise level of the MMB survey was $\le$0.2\,Jy.
This is similar to the deepest previous unbiased surveys which have had
1$\sigma$ sensitivities of between 0.09\,Jy and 1\,Jy (see \citealt{pestalozzi05} and \citealt{pandian07a} for details). 
From the estimates of \citet{walt05} a 3$\sigma$
threshold at $<$1\,Jy should detect between about 75 and 90 per cent of the
total methanol maser population.

Previous surveys with single dish telescopes, such as that of
\citet{caswell95a}, have shown that often the emission from several
different methanol sources can be blended within the beam.  Additionally,
the uncertainties in the positions derived from single dish observations
at these frequencies are too large to make effective associations and to
identify counterparts at other wavelengths.  Therefore an integral
component of the MMB survey has been to determine the positions of the
masers detected in the single dish survey with sub-arcsecond accuracy
using interferometer observations.

For the Parkes detections which had not previously been observed at higher
angular resolution, we are observing them with the Australia 
Telescope
Compact Array (ATCA), and where possible the Multi Element Radio Linked
Interferometer Network (MERLIN);  the latter, in addition to its ability to 
observe sources near declination zero and further north, has the advantage of even 
higher resolution ($\sim$15\,mas rms position 
uncertainty when taking into account the systematic error of the phase 
calibration together with the individual relative position errors).  

\subsection{A Parallel Survey for Excited OH Masers}

The 1-GHz bandwidth of the MMB receiver provided the opportunity to survey
the Galactic plane in an excited transition of OH simultaneously with the
methanol survey.  Often seen in association with 6668-MHz methanol, the
6035-MHz ($^{2}\Pi_{3/2}$ J = 5/2, F = 3$-$3) excited-state OH maser was
first detected in the late-1960s / early-1970s \citep[e.g.][]{yen69,
rydbeck70,
  zuckerman72, knowles73}. After a number of small-scale targeted
observations \citep{knowles76, guilloteau84}, an extensive search by
\citet{caswell95c} detected 72 sources (of which 52 were new to that
survey).  Currently $\sim100$ sources are known
\citep{caswell97,caswell03}.  This maser's susceptibility to Zeeman
splitting can provide additional insight into the magnetic fields present
at sites of high-mass star formation.

\subsection{The Methanol Multibeam Receiver} 
\begin{figure} \centering \includegraphics[width=6cm]{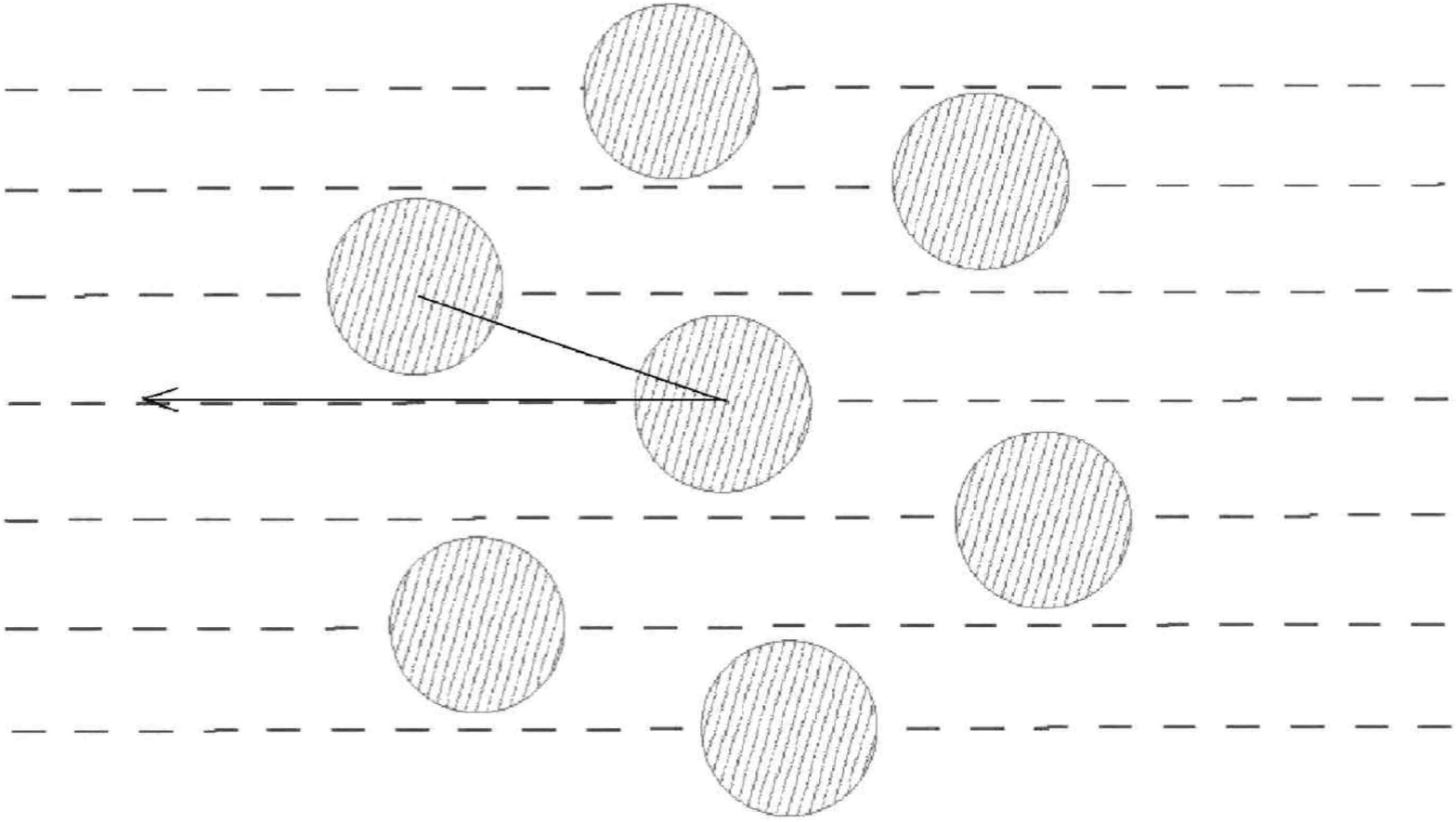}
\caption{\small 
The footprint of the MMB receiver showing the 19.1$^{\circ}$ orientation of the 
array used to map the Galactic plane.  For the Parkes telescope observations the footprint spans a diameter of approximately 15
arcmin.
} \label{BeamPat} \end{figure}

To make the MMB survey feasible required a new purpose-built 
multibeam receiver system, prompting construction of a new 
7-beam receiver jointly by Jodrell Bank Observatory (JBO) and
the Australia Telescope National Facility (ATNF).  The receiver
comprises seven dual circular polarization feeds, arranged in a 
hexagonal pattern around the central feed (Fig.\,\ref{BeamPat}).  
It incorporates low noise indium-phosphide High Electron Mobility 
Transistor (HEMT) amplifiers, which provide a 1-GHz bandwidth.
A 500-Hz switched noise diode is available for continuous
calibration in the presence of varying incoming power levels.
The 
receiver system is mounted at prime focus.  Rotation of the receiver 
package is required to track parallalactic angle, necessitating a 
complex arrangement of flexible cabling for the 28 receiver outputs 
and the helium cooling lines.  

\section{Multibeam Observational Systems and Analysis}

\subsection{MMB Survey Techniques}
The complete MMB survey will cover the full Galactic plane, $-180^{\circ}$
$<$ \textit{l} $<$ 180$^{\circ}$ with a latitude range of
$\pm$2$^{\circ}$.  In this paper we focus on the southern Galactic plane 
portion of the survey using the Parkes 64-m telescope;  the planned 
northern hemisphere extension will adopt a similar strategy. For the
Parkes survey region of $-$174$^{\circ}$ $<$ \textit{l} $<$ 60$^{\circ}$,
 the area observed was 
divided into blocks of 2$^{\circ}$ longitude by 4$^{\circ}$ latitude. 
Adjacent blocks have no overlap, and in the final reduction are combined 
seamlessly.  The observing parameters are summarised in 
Table~\ref{paratable}.

For the Parkes MMB receiver configuration, adjacent beams are separated by 
6.46 arcmin.  The FWHM of each beam is 3.2 arcmin at 6668 MHz and 3.4 
arcmin at 6035 MHz. The width of the total multibeam footprint is $\sim$15 
arcmin. For this hexagonal feed configuration, an orientation at an angle of
19.1$^{\circ}$ to the scan direction (the plane of the Galaxy, see Fig.\,\ref{BeamPat}) 
maintains equally spaced tracks for the seven beams 
\citep*{condon89}.

Scans were conducted across 2$^{\circ}$ in
longitude at a rate of 0.1$^{\circ}$ per min, dumping spectra every 5
seconds.  The parallactic angle tracking was updated continuously
throughout scans.  The beams were then displaced by 1.07 arcmin in latitude
before scanning back. There was then a larger displacement of 15 arcmin to
the next scan pair, thus fully sampling a block of Galaxy of 2$^{\circ}$
in longitude by 4$^{\circ}$ in latitude with 32 scans (16 pairs of forward and backward scans). For each block, a
single schedule file for the Parkes Telescope Control System (TCS) was generated
to define the paths of the 32 scans. Only one pass was
needed as there was minimal radio frequency interference (RFI) at 6668 and
6035 MHz at the Parkes site. 
The noise equivalent flux density of the system temperature at both 6668~MHz and 6035~MHz was 60\,Jy. 
Averaged across all the observing sessions,
this varied by less than 5\,Jy between the different 
beams at each polarization and frequency. Across a
2$^{\circ}$ by 4$^{\circ}$ block, the sensitivity of a final cube 
generally varies by no more than $\pm$3.5 per cent (due to significant oversampling), but is of course worse in the 
direction of the few continuum sources that are strong enough to 
significantly increase the system temperature.
The pointing accuracy of the Parkes Telescope is typically better than 15 arcseconds.

\begin{figure*} 
\begin{center}
\includegraphics[width=16.5cm]{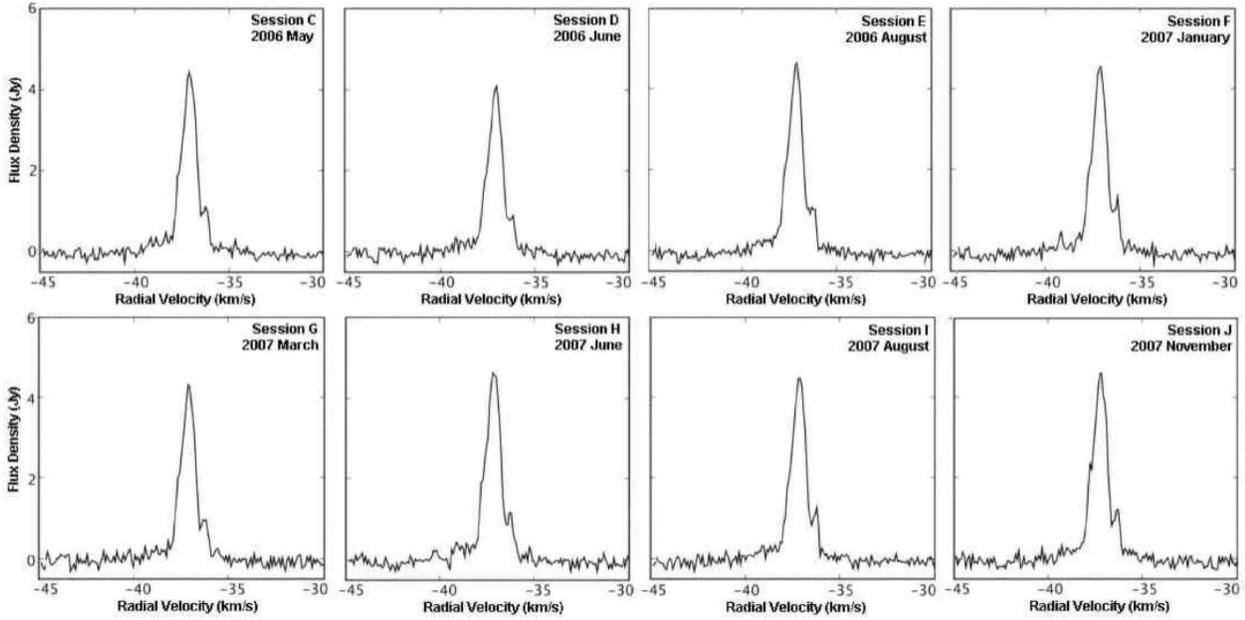}
\end{center}
\caption{6668-MHz methanol source G300.969+1.148 as observed during observing
runs 3 to 10 (for the first two runs a different source was
observed). The median peak flux density is 4.4\,Jy with maximum 
variation $\pm$ 10 per cent.  Intrinsic source variability contributes to the
scatter, but variability within a session was negligible.} \label{300p96meth} \end{figure*}

\begin{figure*} 
\begin{center}
\includegraphics[width=16.5cm]{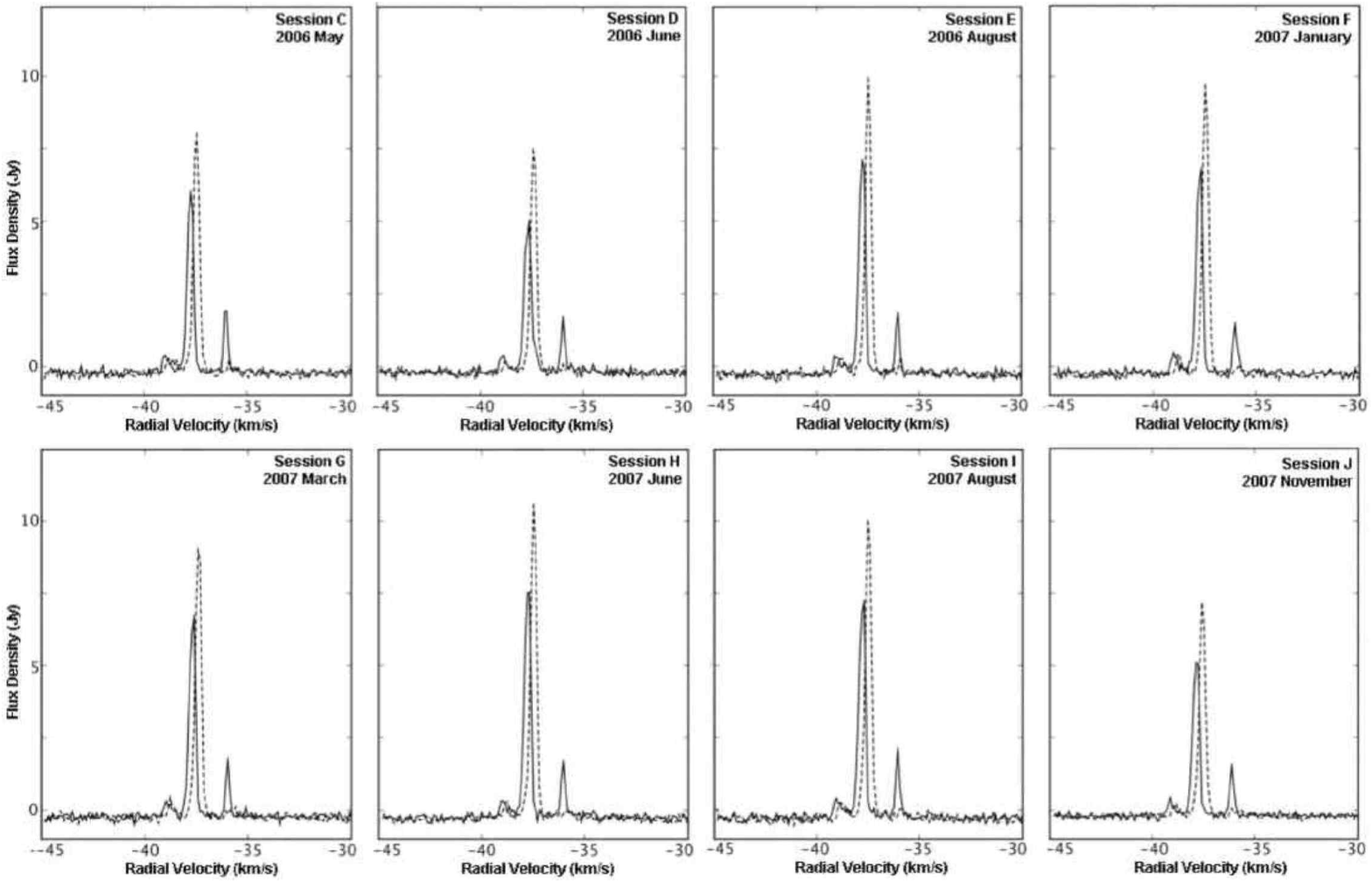} 
\end{center}
\caption{6035-MHz excited-state OH source G300.969+1.148 as observed during observing
runs 3 to 10; the dashed line is the LHC polarization, the solid is the RHC 
polarization. Flux
density scale refers to a single polarization. Over the period from 2006 
May to 2007 November, the LHCP had a median peak flux density of 9.0\,Jy and the 
RHCP 6.4\,Jy.  The range of variation was $\pm$ 20 per cent (with sessions 
D and J showing the largest variation from the mean).  Much of this 
variation is likely to be intrinsic, with the source sometimes twice as strong 
as 5 years earlier \citep{caswell95c}. However, similarly to the methanol,
variability within a session was negligible.} \label{300p96oh} \end{figure*}

The noise diode is used throughout the survey observations for amplitude calibration.
However, its equivalent flux density value is different for each of 
the 28 recorded channels (7 beams, 2 frequencies, 2 polarizations), and to 
measure these values, and also check for pointing offsets, the continuum 
source 1934-638 was observed most days. This used a ``SPOT'' schedule,
which scanned each beam in
turn across the source in RA and Dec. 1934-638 was assumed to have a
flux density of 3.9\,Jy at 6668~MHz and 4.4\,Jy at 6035~MHz.  Calibration stability was further
checked by observing on most days a known maser source using ``MX'' mode,
in which a source was tracked, and the pointing centre cycled through each
of the beams. The most commonly used source was G300.969+1.148 for both 
methanol and excited OH (Fig.\,\ref{300p96meth}
and \ref{300p96oh}). From the repeatability of these
measurements we estimate the flux density calibration to be acccurate
to a few percent.

The first science run of the MMB occurred at Parkes between 22nd and 31st
January 2006. Subsequently a further 101 days spread over the following 22
months (split into ten sessions designated A to J) has yielded a total of
111 days of observations and completed the surveying of the Galactic plane
between $l = -174^{\circ}$ and $l = 60^{\circ}$.

\begin{table} \centering \caption{\small Table of survey parameters for
the southern hemisphere observations. The positional accuracies are listed
as those
  achievable with the ATCA first and where possible MERLIN
second. $^{1}$Noise equivalent flux density calculated from average of all beams and polarizations across all the observing sessions. Based
on observations of 1934-638 and therefore applicable away from the Galactic plane. For the inner Galaxy there can be an addition of several
Jy from continuum emission. $^{2}$Calculated
  as the median of the individual median noise values from each cube.} \begin{tabular}{l c c} \\ \hline
 & \multicolumn{1}{c}{Methanol} & \multicolumn{1}{c}{Hydroxyl}\\ \hline 
Rest Frequency (MHz)  & 6668.519 & 6035.093 \\ 
Beamwidth (arcmin) & 3.2 & 3.4 \\ 
Velocity coverage per setting (km\,s$^{-1}$) & 180 & 200 \\ 
Velocity spacing (km\,s$^{-1}$) & 0.0875 & 0.0971\\ 
Average system noise$^{1}$ (Jy) & 60.1 & 59.9 \\ 
Typical survey rms noise$^{2}$ (Jy) & 0.17 & 0.17\\ 
Positional Accuracy (arcsec) & 0.1/0.01 & 0.1/0.01 \\ 
\end{tabular} 
\label{paratable}
\end{table}

\subsection{Targeted Follow-up}
In addition to the survey scanning observations, all
detections were later observed in pointed observations 
to provide low noise spectra. 
These observations tracked the source,
cycling the pointing centre through each of the beams (adopting the ``MX''
mode). These utilised the positions derived from the high-resolution observations
described in \S \ref{highres}. Each beam was on source for 1 minute, providing an effective
integration time of 7 minutes and a 1$\sigma$ noise of $\sim$0.1\,Jy.

\subsection{Multibeam data acquisition system}\label{dataacq}

Spectra were taken using the multibeam correlator and the wideband
correlator at Parkes, integrated into a single system.  During the first
two observing runs, pending some necessary correlator modifications,
data were not recorded at 6035 MHz for the right hand circular 
polarization (RHCP) of three outer beams (2, 4, and 6).

The observing bandwidth was 4~MHz, which corresponds to a velocity range of
180\,km\,s$^{-1}$ at 6668~MHz (200\,km\,s$^{-1}$ at 6035~MHz ), but 150\,km\,s$^{-1}$ was taken as fully usable (and we incorporated a 10\,km\,s$^{-1}$
overlap for regions with multiple centre velocities). The centre
velocities and range for observations were determined from the
Galactic CO emission of \citet{dame01} (Fig.\,\ref{VELDIST}). To fully cover the required
velocities there was one setting for $|l| > 20^{\circ}$;
two for  $20^{\circ} > |l| > 6^{\circ}$;
three for $6^{\circ} > |l| > 2^{\circ}$; and
finally four for $2^{\circ} > |l|$ .

The two first local oscillators were set to provide Doppler tracking, so
that the centre of each 4-MHz band was kept at a constant radial velocity
with respect to the kinematic Local Standard of Rest (LSR), for both the
methanol and the OH spectral lines. The correlators were configured to
provide 2048 frequency channels across each 4-MHz band, at each line
frequency and each polarization.  Altogether 28 2048-channel spectra were
taken every five seconds in normal survey scans.
The channel spacing of 1950~Hz corresponds to a velocity spacing of
0.09\,km\,s$^{-1}$ at 6668~MHz; this provides an effective velocity
resolution of 0.11\,km\,s$^{-1}$, which is sufficient to resolve the narrow
maser lines (typically 0.2$-$0.3\,km\,s$^{-1}$ in width). The spectra were
not smoothed.

\begin{figure}
 \centering 
 \includegraphics[width=8cm]{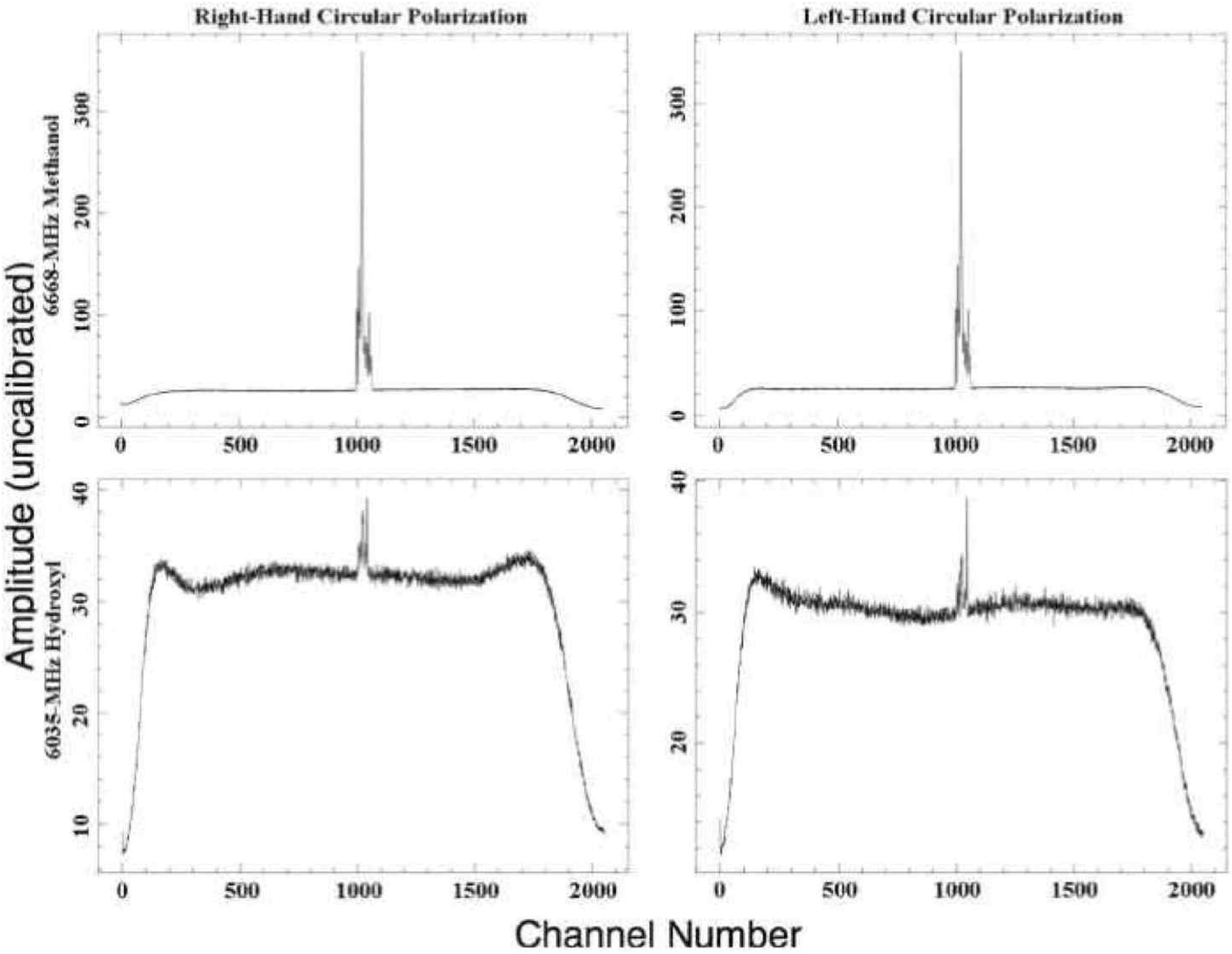} 
 \caption{\small
First light total power spectra of G309.921+0.479 from 2006 January 22.
The upper panels show the 6668-MHz methanol spectra (right and left circular
polarizations) and the lower panels show the 6035-MHz OH. Increasing channel
number corresponds to increasing frequency. Details of the typical bandpass
response can best be seen on the 6035-MHz spectra. }
\label{first_light} \end{figure}

Fig.\,\ref{first_light} shows the first light spectra obtained at Parkes
on 22nd Jan 2006, with the source G309.921+0.479 in the main beam.

\subsection{Multibeam Data Reduction} 
Initial reduction of the survey data was performed with the program 
{\sc  livedata} and its sister program {\sc
gridzilla}, developed by
Mark Calabretta, for processing of multibeam spectral line data\footnote{http://www.atnf.csiro.au/people/mcalabre/livedata.html}.
The capabilities of this package, and its first use (in the HI Parkes All 
Sky Survey, HIPASS), have been described extensively by \citet{barnes01}.  
While some of our 
procedures are the same as those used 
in the HI survey, there are a number of major differences, as we describe 
here.  In particular, we do not encounter interference problems, and have 
much less need for robust median statistics.  

A raw correlator spectrum is a
composite of the maser spectrum, bandpass spectrum, noise and
potentially, baseline ripples.  The bandpass spectrum is the dominant
component and was removed through the division of the correlator spectrum
by an estimate of the bandpass (reference) spectrum.

An individual reference spectrum is required for each polarization of each
frequency for each beam, and is obtained for each 20-minute scan using
{\sc  livedata}. Each reference spectrum is formed from the median of the
240 spectra obtained within a scan. If a maser occurs within a scan, it will
persist for only 15 of the 240 spectra and thus have negligible influence on 
the median. The spectra were not smoothed since we do not have features in 
the spectrum narrower than our resolution (in contrast to the HIPASS 
survey, for which such features caused serious problems).

Following the bandpass correction, the spectra were calibrated for flux density using
the table of values obtained earlier from the noise diode calibration.  A fourth order
polynomial was then fitted to the baseline, and the first and last 100 edge channels
 filtered out.  The fourth order polynomial provided a
15$-$20 per cent improvement of the systematics over other baseline fits,
and was of a low enough order as to not distort any emission or absorption
features that may be present.

\subsection{Gridding and Imaging}

The {\sc  livedata} processing of a 2$^{\circ}$ $\times$ 4$^{\circ}$ block 
of correlator data produced 32 scan files of flux density calibrated and bandpass
corrected spectral data. Each scan consisted of 240 spectra for each 
polarization of each beam. The \textsc{gridzilla} package is used
to convert the individual spectra into position-position-velocity data cubes. The standard
gridding process defines a uniform grid in the observed region and 
determines how the spectra contribute to a pixel through applying weights 
which can be chosen to optimise various parameters, such as the 
signal-to-noise ratio for a point source, the spatial resolution, or minimising
susceptibility to interference.  A smoothing radius was
first chosen to determine which spectra contribute to the pixel.  
The choice of smoothing radius affects the final gridded beamsize and affects the correlation between
pixels and hence the image noise. To balance resolution and positional
accuracy with robustness, a top-hat smoothing kernel with a diameter of 
4.4 arcmin, and hence a radius cut-off of 2.2 arcmin, was used.  
Typically, for a total intensity image, 50 spectra contribute to a pixel,
25 from each of the two polarizations.  The pixel spacing was chosen
to be 1 arcmin so as to be both a convenient value, and somewhat 
oversampled relative to the Nyquist requirements.

To estimate the flux density at a given pixel, the mean of the contributing
spectra was taken, based on the separation of each spectrum from 
the pixel. We normalise the intensity based on the beam response at this 
offset, and then down-weight each value by the square of the beam 
response. 

\begin{figure} \begin{center}
\includegraphics[width=8cm]{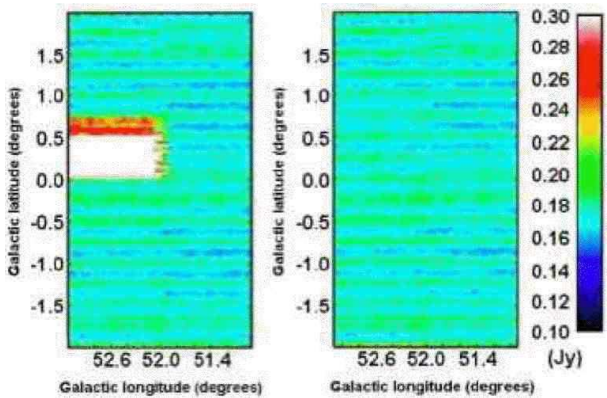} \end{center} \caption{Noise
maps (after flattening the baseline) for a cube made with initial data (left) and final
cube made from repeated scans (right). For the final cube the mean noise level across 24657
pixels in the interior of the map is 0.17\,Jy with a 1-sigma dispersion of
0.01\,Jy.  The minimum value is 0.13\,Jy and the maximum 0.24\,Jy and the
colour scale (online version) is set to span this full range.  The edge effect due to individual scans can be seen in left cube, in the area of high noise.
This causes significant baseline fluctuation, in addition to not fully
sampling and giving potentially erroneous positions. All are removed
through the re-gridding process, as seen on the right, where the noisy data
has been replaced with additional observations.
}
\label{Noise} \end{figure}

Initial cubes were made at the completion of each block (i.e. for the 32 
scans a 2$^{\circ}$ $\times$ 4$^{\circ}$ image
plane is constructed);  this was adequate for initial source detection
(section \ref{detect}), but sources near the edge of a block 
cannot be measured precisely without the additional information from 
the adjacent block. Since there is no overlap between adjacent observing  
blocks, in our final processing we choose to generate cubes of 
2.2$^{\circ}$ $\times$ 4$^{\circ}$, centred at the block junctions so as 
to seamlessly 
blend the adjacent blocks.  The increased 2.2 degree longitude span of 
these final cubes gave a small region of overlap between adjacent 
cubes which ensured that the analysis on any source 
could be made without any edge effects.  The velocity was also
cleanly limited when cubes of differing centre velocities were combined
(Fig.\,\ref{Noise}).

\begin{figure} 
\centering 
\includegraphics[width=8cm]{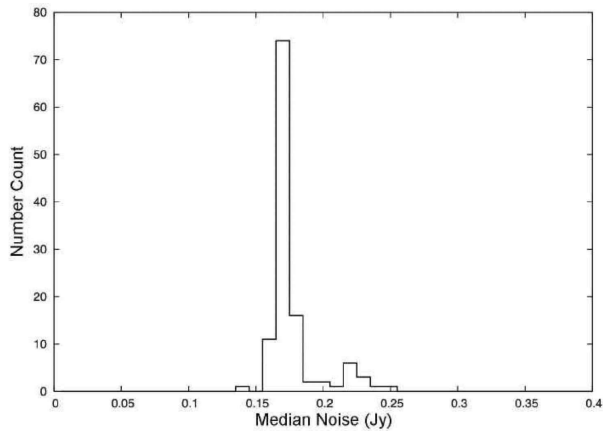} 
\caption{Initial distribution of noise across all the cubes for methanol.  A few
cubes were observed in poor weather, causing the secondary noise peak in
the histogram, and these will be re-observed.} 
\label{Noisedist} 
\end{figure}

For 6668-MHz methanol, data cubes of total intensity (Stokes I) were 
produced, combining the right and left circular polarizations. However, 
for 6035-MHz OH, since the maser emission is typically highly polarized, 
additional separate left hand circular and right hand circular polarized 
emission cubes were produced and examined. The distribution of median 
noise levels across all the final cubes can be seen in Fig.\,\ref{Noisedist}.

\subsection{Source Detection}\label{detect}
Sources were identified in the survey cubes with
an algorithm based on that of the ATNF Spectral line Analysis Package\footnote{http:///www.atnf.csiro.au/computing/software/asap/}.
The
{\sc ASAP} 1-dimension routine was run for each of the $\sim$32000 spatial pixels 
contained within a 2.2$^{\circ}$ $\times$ 4$^{\circ}$ Galactic data cube.

The algorithm operates initially on the spectrum at each position.
The value in each spectral pixel (channel) is compared to a baseline
estimated from a linear fit to the intensity as a function of frequency in a spectral
window covering 5 per cent of the bandwidth ($\sim100$ channels) centred on
the channel under examination. This window was narrower than any
baseline fluctuations, but significantly broader than a typical maser
line.  To classify as a detection, three consecutive channels must have
values more than 4.7 times the root mean square deviation of
their respective baseline fits (i.e. $4.7/\sqrt{3}$ per channel).
This threshold was found to be sufficiently high to limit the number
of false positives (noise above the threshold), while detecting the
majority of real sources.

The algorithm then searched for the tails of emission by extending in
both directions in the spectral domain until the difference between
the channel flux density and the linear baseline fit changed sign. Spectral
lines wholly located in the first 75 (lowest frequency) and last 150 (highest frequency) channels of each
spectrum were excluded as they were likely to be due to the edge
effects of the bandpass. Once a line was detected, the channels
containing it were excluded and the baseline was recalculated. Once
the end of the spectrum was reached the process started again,
iterating until no further lines were found. Several lines could be
detected in one pass, but several iterations were required to detect
weak lines in the presence of strong lines. The noise level was
determined at each iteration of the algorithm through calculating the
average of the variances across the spectrum, excluding 20 per
cent of the largest absolute values as well as channels which met the
detection criteria. 

To search for broader emission the whole process was repeated taking
the average of each pair of channels, adopting a smaller sample box of
50 channels and reducing the signal-to-noise threshold by $\sqrt2$.

When a detection was made in a pixel, all neighbouring spatial pixels were tested
in the same way, with successive testing of neighbouring pixels until the signal
dropped below the threshold, or the image edge was reached. The channels for
each pixel belonging to the new source were merged, but if more than one
distinct spectral line component was found, then the spatial separation 
was compared. If the components were separated by more than one pixel (1 arcmin) the
source was split into two (or more) appropriate positions.  If sources were
close both spatially and spectrally they were combined into a single
source.  However, there were only a few sources that were falsely combined 
and these were easily visually separated on inspection of the cubes.

Once this process had iterated across every pixel in a cube, a source list and a 
series of spectra were compiled.  The source list was verified by eye, 
discarding spurious detections (typically from the edge of the bandpass), 
and reassessing the velocity range of each source. The resultant maser 
positions were accurate enough for follow-up observations with the ATCA.
The performance of the algorithm was tested with a cube of Gaussian 
noise and agreed well with the number of false positives expected for the 
chosen signal-to-noise threshold. The completeness of the detection process
is discussed in detail in section \ref{sensicompi}.

\subsection{Targeted Observations Data Reduction}
\begin{figure}
 \centering \includegraphics[width=8cm]{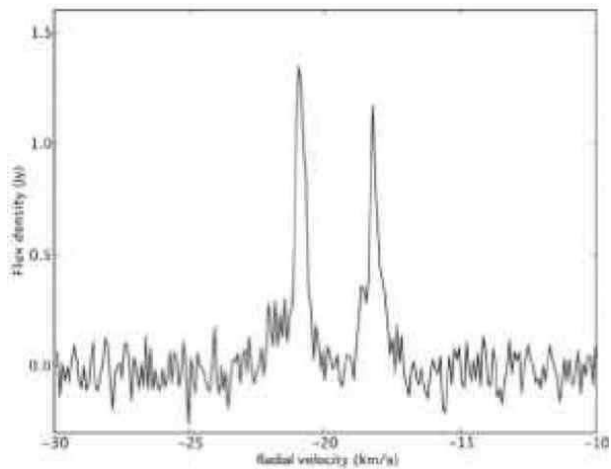} \caption{\small
An example of an outer galaxy maser found by the MMB
survey, G027.000-0.041. It has kinematic Galactocentric and heliocentric distances of $\sim$12\,kpc and $\sim$18\,kpc respectively.}
\label{OuterExample} \end{figure}

The data from the targeted observations employing the `MX' schedule were
processed using {\sc asap}. The bandpass reference for
each beam was estimated using the median spectrum of the six `off-source'
positions, and each median reference then combined with the corresponding
total power spectrum `on-source' to determine a baseline- and
gain-corrected quotient spectrum. For methanol the polarizations were averaged
together.  The seven spectra were then combined to
give a final spectrum with an effective 7 min integration time and a
typical rms noise of less than 0.10\,Jy in the total intensity spectrum.  
Flux
densities were calibrated from the noise diode data and thus ultimately 
linked to the `SPOT' observations of the
continuum source 1934-638. 
The spectrum of a new detection from the survey is shown in Fig.\,\ref{OuterExample}.
The source lies in the first Galactic quadrant and its negative velocity shows that it is a far-side outer Galaxy source.

\section{Accurate Position Measurements}\label{highres} 
All southern sources detected in the Parkes MMB survey without a previously confirmed 
high resolution position were observed with the ATCA. 
Northern sources, and those near declination zero, were observed with MERLIN.

The ATCA observations were taken: 2006
March/April and December;  2007 February, May, July and November; and 2008
January and August. 
The observations were obtained with 6-km E$-$W array configurations, and typically consisted of a loop of targets plus a nearby 
phase calibrator repeated several times within a $\sim$10 hour period. The
time on source was typically a few minutes, but was increased to more than
1 hour for weak targets. The correlator was configured with a 
4-MHz bandwidth (as for the Parkes observations), but with 1024 frequency
channels (rather than 2048), so as to allow a simultaneous search for 
continuum emission at 8.64 GHz.  The primary ATCA flux density calibrator 1934-638 was observed 
in each session, and usually 
1921-293 as a bandpass calibrator. Processing was performed 
using the packages {\sc aips} or {\sc miriad}.  
Visibility data for each source were calibrated and then converted into a spectral cube
image and cleaned.  The typical rms noise is 40-60\,mJy per 3.9 kHz channel and 
positions are accurate to 0.1 arcsec.

The MERLIN observations were taken between 2006
April and 2007 November. Observations were taken in phase referencing mode with 5/6 cuts
of 10 minutes on the source, at intervals of $\sim$2 hours spread over an
$\sim$10 hour period.  Each cut of a source was preceded and followed by $\sim$2 minutes on the
phase calibrator. The observations used five of the six MERLIN
telescopes (the MKII, Darnhall, Tabley, Knockin and Cambridge), with a 
longest baseline of 217 km, which gives a synthesized beam size
of 43 mas at 6.7 GHz. The narrowband correlator setup was 1-MHz bandwidth
and 512 channels and the wideband setup was 16-MHz bandwidth with 16
channels. The MERLIN data were initially reduced with local MERLIN software
and converted to FITS format with preliminary bandpass scaling, then all
further processing was with
{\sc aips} \citep{merlin03}. The primary flux density
and bandpass calibrator was 3C84 (and on occasion 3C286). This was observed in
both wide and narrowband configurations.  The phase calibrator, chosen
depending on the sources, was observed purely in the wideband setup for
optimal signal-to-noise. Sources were observed at fixed frequency.  Flux density and phase
were calibrated in the narrowband and wideband data respectively, and then the solutions
applied to the source data. Inspection of the spectra allowed the brightest channel 
to be selected and imaged to obtain an accurate position. 
The typical rms noise is 25\,mJy per channel and the positions
are accurate to $\sim$15\,mas. MERLIN observations were made of not only the new detections
in the 20$^{\circ}$ to 60$^{\circ}$ longitude region, but also the known sources from the survey of \citet{szymczak02}
which had no published high-resolution observations. 

\section{Parallel Surveys} 
\subsection{Magellanic Clouds Survey} 
In addition to the Galactic plane survey, we have also conducted the first
complete survey of the Large and Small Magellanic Clouds for 6668-MHz
methanol and 6035-MHz excited-state hydroxyl masers \citep{green08a}. This
included higher-sensitivity targeted searches towards known star-formation
regions. The observations yielded the discovery of a fourth 6668-MHz
methanol maser in the Large Magellanic Cloud (LMC), found towards the
star-forming region N160a, and a second 6035-MHz excited-state hydroxyl
maser, found towards N157a.  We also re-observed the three previously
known 6668-MHz methanol masers and the single 6035-MHz hydroxyl maser. All
observations were initially made using the MMB receiver on the
Parkes telescope when the Galactic plane was not visible, and accurate positions were
measured with the ATCA.  \citet{green08a} present an analysis of the water and
ground-state OH maser populations in addition to the excited-state OH and
methanol, and found the LMC maser populations to
be smaller than their Milky Way counterparts by up to a factor of
$\sim$45. We failed to detect emission from either transition in the Small
Magellanic Cloud.

\subsection{Pulsar Piggyback Programme}\label{PuPiPr}
\begin{figure} \begin{center}
\includegraphics[width=8cm]{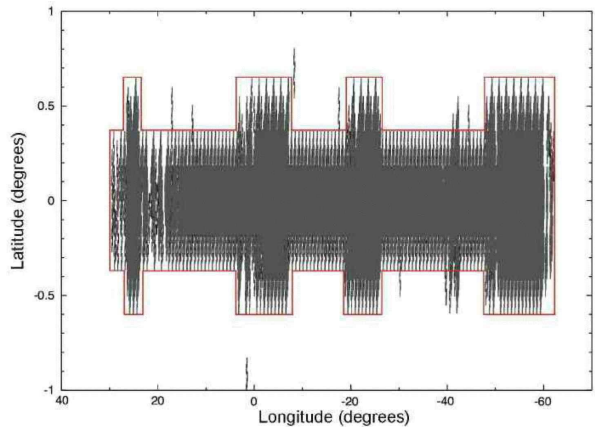} \end{center} 
\caption{\small Regions covered with the deep pointed integrations of the Piggyback survey.
Shading represents actual area covered by beams.
Solid lines outline the main region surveyed ($\sim$85\,degree$^{2}$),
of which 60 per cent is covered by the beams ($\sim$51\,degree$^{2}$).}
\label{PigCoverage} \end{figure}

The 6-GHz multibeam receiver was also used for a Galactic pulsar
search (Parkes project P512 conducted by Johnston and colleagues; \citet{obrien08}) and we were able to `piggyback' spectral line observations for
the 6668-MHz methanol transitions on this program.  The pulsar
observations were taken at two frequencies, one of which included the
6668-MHz methanol frequency.  Piggyback observations occurred over
eight sessions between February 2006 and August 2007.  The
requirements of the primary pulsar observations prevented Doppler
tracking and switching of the noise diode during the piggybacked
spectral line observations.  Due to these restrictions the methanol
observations were made using a single fixed centre frequency (6668.5
MHz) that was suitable for a wide range of Galactic positions and a
bandwidth of 8 MHz, with 4096 spectral channels.  The piggyback survey
thus has the same spectral resolution as the main survey, but double
the velocity coverage and only observed the methanol transition and
not excited OH. The pulsar observations were made using long pointed
integrations (as opposed to the scanning strategy employed for the MMB
survey).  The observations primarily targeted the Galactic centre, covering
an area of $\sim$51\,degree$^{2}$ distributed over $\sim$85\,degree$^{2}$ on 
the sky (Fig.\,\ref{PigCoverage}).  The typical integration times were 17 min,
but a small region near the Galactic centre was covered by even
deeper, 5-hour, integrations.

The different observing strategy of the piggyback observations
(compared to the main MMB survey) necessitated the development of a
separate data reduction approach.  For the piggyback observations the
data were processed using {\sc asap}.  As noise diode switching was
disabled for the pulsar observations there were no system temperature
measurements available for the piggyback data.  We assumed a system
noise equivalent flux density of 65\,Jy for each beam in all observations, approximately
8 per cent higher than cold sky values
(see section 3.1).  We chose this higher value partly to be
conservative in our sensitivity estimates and partly to account for
the fact that the pulsar observations were more concentrated to
regions of higher than average Galactic background continuum emission.  The
general reduction strategy in the case of the 17 min integration data
sets was to form a single reference spectrum for each polarization
and beam for each day.  The reference was formed using all the
spectra for that polarization and beam on a particular day and finding
the median value for each channel in the spectrum.  The quotient
spectra that result from these median references contain baseline
ripples with amplitudes of around 1\,Jy which cannot be removed even
with a high order ($>$ 15th) polynomial.  We found that a running median
filter (of width 50 channels) produced a smoothed copy of the
spectrum, which when subtracted from the original was very effective
at removing the baseline ripple with minimal effect on real maser
emission.  For strong masers (peak flux density $>$ 10\,Jy) with a
velocity range comparable to the filter width this method results in a
negative dip in the baseline around the maser emission.  This does not
effect our ability to detect these masers (all of which are found in
the main MMB survey anyway) and the weak (and typically narrow
velocity range) masers that are the primary target of the piggyback
survey do not suffer this problem.  The piggyback survey utilises the
same line-finder algorithm employed in the main survey, but using a
lower detection threshold of two consecutive channels exceeding three times the rms.

\begin{figure}
 \centering \includegraphics[width=8cm]{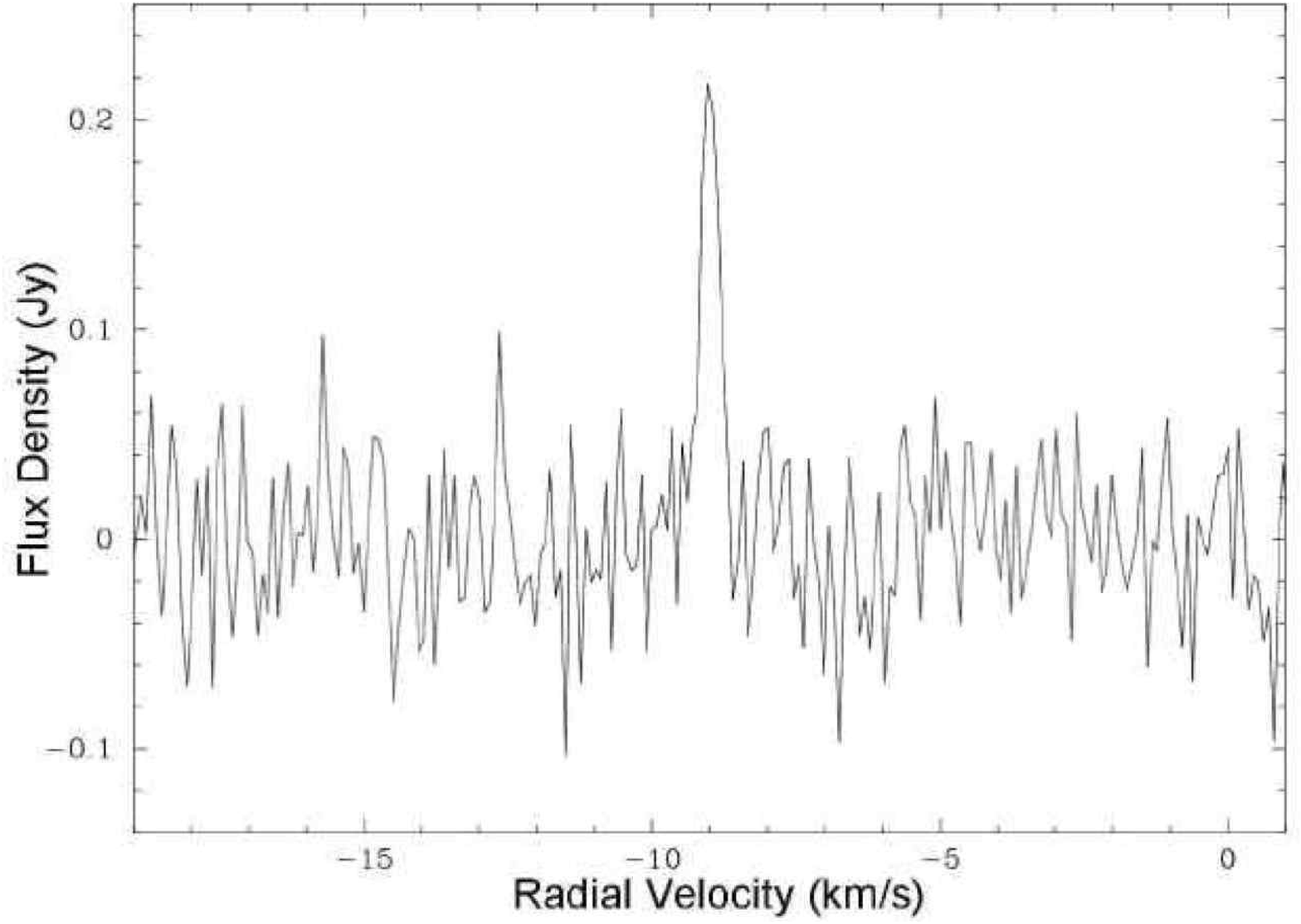} \caption{\small Example
spectrum from the Piggyback data showing source
G001.40+0.09. Note the low rms noise level of 0.042\,Jy; this weak maser would 
not have been detected by the main survey.} \label{PigExample} 
\end{figure}

The effective per-beam integration time for the main MMB survey is approximately one
minute and so we would expect that the piggyback survey with
integration times of 17 minutes should be a factor of approximately four
more sensitive.  This is indeed the case, with the typical measured
rms for a 17 minute integration being 42\,mJy.  This is approximately a factor
of two more sensitive than the Arecibo survey of Pandian et al. (2007).  Thus, the piggyback survey is the
most sensitive large area methanol maser survey that has been made to
date.  These data provide a much deeper, but undersampled survey,
which will be used to probe the low flux density end of the maser
population and the variability of these sources. Fig.\,\ref{PigExample}
shows an example spectrum of a weak new source from the piggyback
observations; it demonstrates that the piggyback
observations are capable of detecting masers with peak flux densities
as low as 0.2\,Jy and this is confirmed through numerous detections of
strong masers in distant sidelobes.  The sensitivity of the piggyback
search is such that masers with peak flux densities in excess of a few
hundred Jy can be weakly detected in pointings more than half
a degree from their true location.  A preliminary analysis of
approximately 25 per cent of the piggyback data ($\ell \le 10^{\circ}$)
suggests of the order of 40 new detections, the majority below 0.5\,Jy.

\section{Survey Sensitivity \& Completeness}\label{sensicompi}
\subsection{Efficiency of Maser Detection} 
In order to test the efficiency and accuracy of the
detection methods employed by the MMB, we constructed several test cubes
containing simulated masers. The process involved creating a skymap of
positions of masers together with associated data files containing
velocities and flux densities for each maser position. To create realistic simulated
maser sources we took some observed 10\,Jy masers and rescaled them to 
represent typical sources at a range of flux densities. The 10\,Jy masers 
were selected to have features across a realistic
distribution of velocity ranges and were given randomly distributed peak
velocities. The rescaled peak flux densities were 0.5, 0.6, 0.7, 0.8, 0.9, 1.0 and 1.1\,Jy. 
Forty simulated masers were created at each flux density level, and split into
two batches with random longitudes and latitudes between the ranges of 220$^{\circ}$
$-$ 222$^{\circ}$ and $-$2$^{\circ}$ to $+$2$^{\circ}$ respectively. This range was chosen so that the simulated
masers could be added to an empty real data cube. The simulated sources
were then added to the bandpass calibrated data files ({\sc sdfits}).  
This was done by calculating the angular separation and parallactic angle
between the Parkes beam and the source position, then obtaining the
attenuation factor. The simulated spectrum was then scaled using this
factor. The scaled spectra were added to the data files. By adding
the masers to the data files for a real cube, rather than simulated pure
noise, we ensured that the effect of any residual baseline wobble would be
taken into account. The beam model was approximated by varying the model of \citet{james87} to a dish diameter of 64\,m and
allowing for three feed leg shadows. This never the less had
the limitation that it did not fully describe the sidelobe
properties.

\begin{figure} \begin{center} \includegraphics[width=8cm]{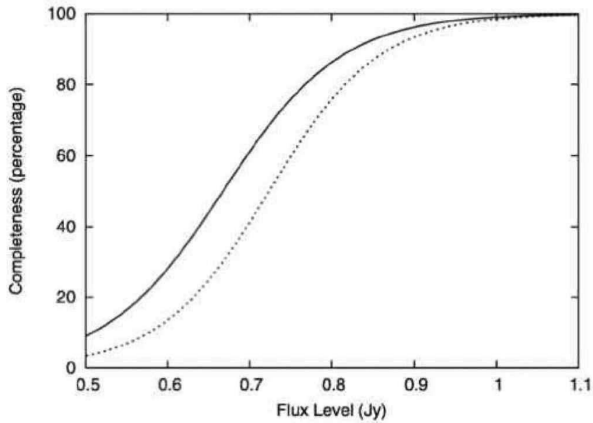}
\end{center} \caption{\small Results of simulation to determine detection efficiency percentage as a function of model
flux density. Lines show the fits of the `algorithm detections' (solid) and the final `catalogued sources' (dotted) to sigmoid functions.
} \label{completeness} \end{figure}

The 14 simulated cubes (with a batch of 20 masers at a given flux density in each
cube) were processed in the same manner as the survey data (section 3.7). This allowed
us to form two sets of results: the detections made by the program (`algorithm detections' in Table \ref{statstable} and Fig.\,\ref{completeness}); and
the subset of these that were then deemed real by eye (`catalogued sources' in Table \ref{statstable} and Fig.\,\ref{completeness}). 
The  inspection by eye removes any false detections by the algorithm.
From the underlying trend we estimate the completeness to be $\sim$80
 per cent at 0.8\,Jy, approaching 100 per cent at 1\,Jy.

\begin{table*} \centering \caption{\small Comparison of the number of
simulated sources versus those detected by the algorithm. The simulated
source cube had a rms noise of 0.18\,Jy and 40 sources were simulated at each flux density
level.
The first column is the input peak flux density of the simulated masers. The second column 
is the proportion of the population of simulated masers which would be expected to fall below the 
detection threshold of 3$\sigma$ (0.54\,Jy) based on a Gaussian distribution of flux densities
about the input level. The third column is the number of detections made by the automated detection algorithm.
The fourth column is the percentage completeness, with the associated 
$\sqrt{N}$ statistical error (again based on a Gaussian distribution). The fifth column is 
the number of sources deemed `real' by visual inspection and the final is the completeness of this.} 
\begin{tabular}{c c c c c c} \\ \hline
Input & Percentage  & \multicolumn{1}{c}{Algorithm} & \multicolumn{1}{c}{Completeness} & \multicolumn{1}{c}{Catalogued} & \multicolumn{1}{c}{Total}\\
Flux & of sources & \multicolumn{1}{c}{Detections} & \multicolumn{1}{c}{of Algorithm} & \multicolumn{1}{c}{Sources} & \multicolumn{1}{c}{Completeness}\\
(Jy) & below threshold &  & \multicolumn{1}{c}{(per cent)} &  & \multicolumn{1}{c}{(per cent)}\\ 
\hline
0.5 & 58.8 & 3 & 8$\pm$4 & 0 & 0\\
0.6 & 36.9 & 11 & 28$\pm$8 & 7 & 18$\pm$7\\ 
0.7 & 18.7 & 25 & 63$\pm$13 & 14 & 35$\pm$9\\ 
0.8 & 7.4 & 35 & 88$\pm$15 & 34 & 85$\pm$15\\ 
0.9 & 2.3 & 37 & 93$\pm$15 & 36 & 90$\pm$15\\ 
1.0 & $\le$0.1 & 36 & 90$\pm$15 & 32 & 80$\pm$14\\ 
1.1 & $\le$0.1 & 40 & 100$\pm$16 & 36 & 90$\pm$15\\ 
\end{tabular} 
\label{statstable} 
\end{table*}

\subsection{Completeness of Catalogue} 
The recent survey with the Arecibo
telescope by \citet{pandian07a} surveyed the region 35.2$^{\circ}$ $\le$
$l$ $\le$ 53.7$^{\circ}$, $|b|$ $\le$ 0.41$^{\circ}$ with a survey threshold of 0.27\,Jy (more than 95 per cent probability
of detection at 3$\sigma$).
The MMB survey region fully encompasses the Arecibo
survey region, and of the 86 detections found by Pandian et al. the MMB
detected 73. The strongest of the non-detections is G037.77-0.22 with peak flux
density 0.82\,Jy, and the remaining 12 had peak flux densities (at the epoch of the
Arecibo observations) between 0.13 and 0.5\,Jy.
We note
the Arecibo survey had slightly narrower velocity channels, $\sim$0.07 km
s$^{-1}$, but was Hanning smoothed. The velocity range was $-$70 to 110\,km\,s$^{-1}$, whilst the
MMB's velocity coverage was:  $-$35 to 115\,km\,s$^{-1}$ for 35.2$^{\circ}$ $< l <$ 40$^{\circ}$; $-$45 to
105\,km\,s$^{-1}$ for 40$^{\circ}$ $< l <$ 46$^{\circ}$; $-$55 to 95\,km\,s$^{-1}$ for 46$^{\circ}$ $< l <$ 50$^{\circ}$;
 and $-$65 to 85\,km\,s$^{-1}$ for 50$^{\circ}$ $< l <$ 53.7$^{\circ}$.  None of the Arecibo sources fall 
outside the MMB velocity coverage.  
Comparisons between the two surveys are affected by the
variability of sources from the time of the Arecibo survey to the MMB 
survey (up to three years for some sources).
Some methanol masers exhibit large variations within less than a year 
\citep*[e.g.][]{caswell95d, macleod96, goedhart04}, and some of the weaker 
Arecibo masers may have decreased in flux density below our threshold.
G037.77-0.22 seems to be an example, since it had a recorded peak flux
density of 0.82\,Jy in 2005 September, but was not detected above 0.4\,Jy by the MMB
survey in 2007 September.
 
Another test of the completeness of the catalogue is obtained by taking
deeper observations to much lower detection thresholds. 
As described in section
5.2, we are able to achieve these through `piggybacking' on pulsar
observations.  These observations will provide a better estimate of the
luminosity distribution within the region and reveal what fraction
of the total population we are detecting with the main survey. The results
of this work will be fully described in a subsequent paper (Ellingsen et
al. in preparation).

\citet*{pandian07b} in a follow up paper to the Arecibo survey used the \citet{walt05}
model distribution to estimate that in the longitude range 40 to 50 degrees,
the fraction of total masers in the Galaxy above the threshold of 0.27\,Jy
would be between 89 and 93 per cent. If the \citet{walt05} predictions
are applied to MMB survey sensitivity we find between 79 and 86 per cent
are above the threshold. A full analysis of the completeness by longitude will be 
discussed in future papers.

\section{Summary} 
The Methanol Multibeam survey, employing a new purpose-built 
7-beam receiver, has completed observing the longitude range 
$-$174$^{\circ}$  $<$  $l$ $<$ $+$60$^{\circ}$ with the Parkes 
radio telescope. More than 800 6668-MHz methanol masers have 
been found to date (including several far-side outer Galaxy masers), 
suggesting a total Galactic population in excess 
of 1000. We have also surveyed the Magellanic Clouds 
\citep{green08a} and have conducted a deeper survey of a much 
smaller region in our `piggyback' observations (Ellingsen et al. in 
preparation). The results of the methanol survey will be released 
sequentially, starting with the Galactic centre, and will incorporate 
accurate positions and spectra for each maser. The excited-state OH 
results will then be released and the data sets will be made available 
online in due course. Follow-up observational programmes have 
already been initiated to probe the maser surroundings.

Through incorporating new technology, multiple radio astronomy instruments, 
and international collaboration, the MMB survey is well on its way to producing 
the largest unbiased, homogeneous Galactic plane catalogues of 6668-MHz 
methanol and 6035-MHz hydroxyl masers to date. The Galactic distribution of 
methanol masers, when combined with distance indicators such as HI self-absorption 
and maser parallax studies, will allow for the determination of the distribution of 
high-mass star formation regions, and thus will have implications for understanding spiral arm 
structure and Galactic dynamics. The precise source positions in the catalogues, when
combined with comparable high-precision observations at sub-mm and far-IR 
wavelengths from future instruments, will hold the key to improving our 
understanding of the conditions that give birth to high-mass stars.

\section*{Acknowledgments} JAG, AA, JCox and DW-McS acknowledge the support of
a Science and Technology Facilities Council (STFC)  studentship. LQ
acknowledges the support of the EU Framework 6 Marie Curie Early Stage
Training programme under contract number MEST-CT-2005-19669 ``ESTRELA''.  
MERLIN is a national facility operated by the University of Manchester on
behalf of STFC. The Parkes Observatory and the Australia Telescope Compact
Array are part of the Australia Telescope which is funded by the
Commonwealth of Australia for operation as a National Facility managed by
CSIRO. The authors greatly thank the engineering staff both at the
Australia Telescope National Facility and Jodrell Bank Observatory, and
the staff at the Parkes observatory for ensuring the smooth running of the
observations. The authors dedicate this paper to the memory of R. J.
Cohen.

\bibliographystyle{mn2e} \bibliography{UberRef}

\label{lastpage}

\end{document}